\documentclass[12pt,english,preprint]{revtex4-1}
\usepackage{times}
\usepackage{array}
\usepackage{longtable}
\usepackage{float}
\usepackage{amsmath}
\usepackage{bm}
\usepackage{graphicx}
\usepackage{amssymb}
\usepackage{color}
\usepackage[FIGTOPCAP]{subfigure}

\makeatletter
\usepackage{fancyhdr}
\usepackage{babel}
\makeatother
\begin{document}

\title{A Physics-Informed Deep Learning Description of Knudsen Layer Reactivity Reduction}

\author{Christopher J. McDevitt}
\email{cmcdevitt@ufl.edu}
\affiliation{Nuclear Engineering Program, Department of Materials Science and Engineering, University of Florida, Gainesville, FL 32611, United States of America}
\author{Xian-Zhu Tang}
\affiliation{Theoretical Division, Los Alamos National Laboratory, Los Alamos, NM 87545, United States of America}

\date{\today}

\begin{abstract}

A physics-informed neural network (PINN) is used to evaluate the fast ion distribution in the hot spot of an inertial confinement fusion target. The use of tailored input and output layers to the neural network is shown to enable a PINN to learn the parametric solution to the Vlasov-Fokker-Planck equation in the absence of any synthetic or experimental data. As an explicit demonstration of the approach, the specific problem of Knudsen layer fusion yield reduction is treated. Here, predictions from the Vlasov-Fokker-Planck PINN are used to provide a non-perturbative solution of the fast ion tail in the vicinity of the hot spot thus allowing the spatial profile of the fusion reactivity to be evaluated for a range of collisionalities and hot spot conditions. Excellent agreement is found between the predictions of the Vlasov-Fokker-Planck PINN and results from traditional numerical solvers with respect to both the energy and spatial distribution of fast ions and the fusion reactivity profile demonstrating that the Vlasov-Fokker-Planck PINN provides an accurate and efficient means of determining the impact of Knudsen layer yield reduction across a broad range of plasma conditions.

\end{abstract}

\maketitle


\section{Introduction}

Inertial confinement fusion (ICF) experiments generate plasmas that encompass an exceptionally broad range of densities and temperatures. This broad range of plasma conditions creates a challenge with regard to selecting the level of physics fidelity required for accurately describing an ICF implosion. In particular, while radiation-hydrodynamic codes have emerged as the backbone to integrated simulations of ICF capsules~\cite{clark2011short, johnson20203d}, such a framework is based on perturbative closures that are only valid for asymptotically small values of the Knudsen number $K_n \equiv \lambda_{mfp}/\Delta$, where $\lambda_{mfp}$ is the particle's mean-free-path and $\Delta$ is the smallest gradient length scale. While the limit $K_n \ll 1$ is well satisfied across a range of ICF conditions, this limit is strongly violated in low density exploding pusher experiments~\cite{rosenberg2014investigation, rinderknecht2015ion}, and due to the strong temperature dependence of the Knudsen number ($K_n \propto T^2$), becomes increasingly suspect for burning plasmas capable of achieving high ion and electron temperatures~\cite{hartouni2023evidence}.

The presence of low to modest levels of collisionality allow for 
strong deviations from a local Maxwellian distribution to develop that cannot be treated by perturbative closures~\cite{chapman1990mathematical, braginskii1958transport}. Such deviations of an ion or electron distribution from a local Maxwellian impact closure quantities such as the heat flux~\cite{spitzer1953transport, epperlein1991practical, schurtz2000nonlocal, chapman2021preliminary, miniati2022learning, mitchell2024reduced}, the magnitude and spatial profile of the fusion reactivity~\cite{henderson1974burn, petschek1979influence, molvig2012knudsen, mcdevitt2014calculation, kagan2015self}, atomic physics rates~\cite{garland2022understanding}, and thus can strongly impact the trajectory of an ICF implosion~\cite{rinderknecht2018kinetic}. In addition, sufficiently strong deviations from a Maxwellian impact diagnostics based on neutron~\cite{frenje2020nuclear, crilly2022constraints} or hard X-ray spectra~\cite{kagan2019inference}.
In this paper we describe how physics-informed machine learning methods enable the development of an efficient surrogate model for the tail ion distribution to be rapidly inferred. Such an approach offers a complement to traditional numerical solvers focusing on the solution to the Vlasov-Fokker-Planck (VFP) equation~\cite{larroche2012ion, taitano2015mass}. Here, rather than relying on data, this approach seeks to embed physical constraints into the training of a neural network (NN). In so doing, the quantity of data needed to train the NN can be sharply reduced, or even eliminated.

As an initial study, we demonstrate the ability of this approach to provide a comprehensive description of Knudsen layer reactivity reduction in the hot spot of an ICF target. In particular, a physics-informed neural network (PINN) is used to solve the time dependent VFP equation for a geometry with one spatial dimension and two velocity space dimensions (1D-2V) in the absence of any data. It is shown that the VFP PINN is able to learn the parametric dependence of solutions to the Vlasov-Fokker-Planck equation, thus enabling a fast surrogate model of plasma kinetic effects.

The remainder of this paper is organized as follows. Section \ref{sec:PCDL} provides a brief description of physics-informed neural networks, with an emphasis on how they may be used to learn the solution space of parametric PDEs~\cite{sun2020surrogate, McDevitt:hottail:2023}. An overview of the system of equations solved is given in Sec. \ref{sec:ME}, along with a description of how customized input and output layers are included for treating the specific problem of the fast ion distribution in the vicinity of a plasma hot spot. Section \ref{sec:FIS} describes the fast ion solution predicted by the VFP PINN. The fusion yield of an ICF hot spot is described in Sec. \ref{sec:KLRR}, along with a comprehensive description of how hot spot parameters impact Knudsen layer yield reduction. Conclusions and a brief discussion are given in Sec. \ref{sec:CD}.

\section{\label{sec:PCDL}Physics-Constrained Deep Learning}


A primary aim of this paper is the development of a PINN customized to treat the VFP equation. Before doing so, it will be useful to briefly review fundamental aspects of the PINN framework, which has emerged as a prominent example of physics-informed machine learning methods~\cite{lagaris1998artificial, karpatne2017theory, karniadakis2021physics, lusch2018deep, wang2020towards}. The present discussion will only focus on the essential concepts, where the interested reader is referred to Ref. \cite{karniadakis2021physics} and references therein for a more detailed discussion. Here, the underlying strategy is to impose physical constraints into the training of a NN. This can be accomplished either via the use of customized layers in the NN that directly constrain the predictions of the NN (i.e. hard constraints), or via the addition of physical constraints into the loss function (soft constraints). A PINN in its simplest form is focused on the latter strategy, with the loss expressed as~\cite{raissi2019physics}:
\begin{align}
\text{Loss} &= \frac{1}{N_{PDE}} \sum^{N_{PDE}}_i  \mathcal{R}^2 \left( \mathbf{z}_i, t_i; \bm{\lambda}_i \right) + \frac{1}{N_{bdy}} \sum^{N_{bdy}}_i \left[ f_i - f \left( \mathbf{z}_i, t_i; \bm{\lambda}_i \right)\right]^2 \nonumber \\
& + \frac{1}{N_{init}} \sum^{N_{init}}_i \left[ f_i - f \left( \mathbf{z}_i, t=0; \bm{\lambda}_i \right)\right]^2
, \label{eq:PCDL2}
\end{align}
where $f$ represents the field being solved for (the ion distribution
in the present paper), $\mathbf{z}_i$ are phase space coordinates
(energy, pitch and a spatial coordinate), $t_i$ is time,
$\bm{\lambda}_i$ represents parameters of the physical system (Knudsen
number, for example), and $\mathcal{R}\left( \mathbf{z}_i,
t_i; \bm{\lambda}_i \right)$ is the residual of the PDE. The second
and third terms on the right hand side set the boundary
and initial conditions, respectively, whereas the first term enforces the PDE.  An additional term
containing any available data~\cite{mathews2022deep, mcdevitt2024physics} can be added to the loss in
Eq. (\ref{eq:PCDL2}), though no data will be used in the present
study. Our motivation will instead be to demonstrate the ability of
PINNs to accurately learn solutions to the Vlasov-Fokker-Planck
equation across a broad range of parameters $\bm{\lambda}$ in the zero
data limit. 

The use of a vanilla PINN such as Eq. (\ref{eq:PCDL2}) will fail to accurately
evaluate the ion distribution at high energies. This is due to the exponentially small number of tail ions compared to bulk ions, such that the tail ion distribution makes very little contribution to the PDE residual given by the first term in Eq. (\ref{eq:PCDL2}). A primary motivation of the present work will thus be to develop a tailored PINN, with a
loss function calibrated to give appropriate weight to the tail ion distribution and custom input and output layers that ensure physical constraints and symmetries of the system are exactly satisfied.

\section{\label{sec:ME}Model Equations}

\subsection{\label{sec:PM}Physics Model}

For modest Knudsen numbers, ions in the bulk plasma will be nearly Maxwellian, and will thus be well approximated by collisional closures based on perturbative Chapman-Enskog expansions~\cite{chapman1990mathematical}. Due to the mean-free-path of an ion scaling with the square of the ion's energy, however, we anticipate strong deviations from a Maxwellian distribution at high energy. A non-perturbative treatment of such deviations can be achieved by utilizing a test-particle collision operator~\cite{Helander-Sigmar:book}, whereby the fast ion population is evolved under the assumption that collisions between the tail and Maxwellian bulk are dominant compared to tail-tail collisions~\cite{tang2014reduced}. In this limit, 
the VFP equation reduces to~\cite{mcdevitt2014comparative}:
\begin{equation}
\sqrt{\bar{E}}\frac{\partial f_a}{\partial \bar{t}} + \frac{\partial}{\partial \bar{x}} \left( V_x f_a\right) = \frac{\partial}{\partial \bar{E}} \left[ \left( V_E + D_E \frac{\partial}{\partial \bar{E}} \right) f_a\right] + \frac{\partial}{\partial \xi} \left[ \left( V_\xi + D_\xi \frac{\partial}{\partial \xi} \right) f_a \right]
, \label{eq:FIS1}
\end{equation}
where the collisional coefficients are defined by:
\begin{subequations}
\label{eq:ME2}
\begin{equation}
V_x \equiv \xi \bar{E}
, \label{eq:ME2a}
\end{equation}
\begin{equation}
V_E \equiv \overline{\nu}^E_a \bar{T}^{3/2} \left[ 1 + \left( \frac{\bar{E}}{\bar{E}_c} \right)^{3/2}\right] - \bar{{\cal E}}_a \xi \bar{E}
, \label{eq:ME2b}
\end{equation}
\begin{equation}
V_\xi \equiv - \frac{1}{2} \bar{{\cal E}}_a \left( 1 - \xi^2 \right)
, \label{eq:ME2bsub1}
\end{equation}
\begin{equation}
D_E \equiv \overline{\nu}^E_a \bar{T}^{5/2} \left[ 1 + \left( \frac{\bar{E}}{\bar{E}_c} \right)^{3/2}\right]
, \label{eq:ME2c}
\end{equation}
\begin{equation}
D_\xi \equiv \overline{\nu}^{\xi}_a \frac{\bar{T}^{3/2}}{\bar{E}} \left( 1 - \xi^2 \right)
, \label{eq:ME2d}
\end{equation}
\end{subequations}
with the normalizations
\[
\bar{E} \equiv \frac{E}{T^{hs}},\quad \bar{x} \equiv \frac{x}{L},\quad \bar{T} \equiv \frac{T}{T^{hs}},\quad \bar{n} \equiv \frac{n}{n^{hs}}, \quad \bar{t} \equiv \frac{v^{hs}_{Ta} t}{L}
,
\]
\[
\bar{{\cal E}}_a \equiv \frac{e_a {\cal E} L}{T^{hs}} \quad \bar{\nu}^E_a \equiv \frac{\nu^E_a L}{v^{hs}_{Ta}},\quad \bar{\nu}^\xi_a \equiv \frac{\nu^\xi_a L}{v^{hs}_{Ta}},\quad \bar{E}_c \equiv \frac{E_c}{T^{hs}}
.
\]
Here, $T^{hs}$ is the hot spot temperature, $n^{hs}$ is the density at the center of the hot spot, $v^{hs}_{Ta} \equiv \sqrt{2 T^{hs}/m_a}$, we have assumed a slab geometry with one spatial dimension $x$, $L$ is a length scale used to normalize the system size (taken to be approximately half the system size), $E\equiv m_a v^2/2$ is the ion kinetic energy, $\xi \equiv v_x/v$ is the pitch, and $\mathcal{E}$ is the electric field. For non-thermal ions satisfying $v>v_{Ta}$, the dimensionless collision frequencies for a DT plasma are given by:
\begin{align}
&\bar{\nu}^E_a \equiv 2 \frac{L}{\lambda_{mfp}} \sqrt{\bar{T}} \left( \frac{n_d}{n_e} \frac{m_a}{m_d} + \frac{n_t}{n_e} \frac{m_a}{m_t}\right)
, \label{eq:ME2sub1a}
\\
&\bar{\nu}^\xi_a \equiv \frac{1}{4} \bar{\nu}^E_a \frac{m_d}{m_a} \left( 1 + \frac{n_t}{n_d} \right)\left( 1 + \frac{n_t}{n_d} \frac{m_d}{m_t}\right)^{-1}
, \label{eq:ME2sub1b}
\end{align}
where $\lambda_{mfp}$ is the mean free path of a thermal ion. The collisional dependence can be described by the Knudsen number defined by:
\begin{equation}
K_n = \frac{\lambda_{mfp}}{L} =  \frac{4\pi \epsilon^2_0 m^2_a v^4_{Ta}}{n_e e^4 \ln \Lambda L} \equiv K_n^{hs} \frac{\bar{T}^2}{\bar{n}}
, \label{eq:ME2sub3}
\end{equation}
where $\ln \Lambda$ is the Coulomb logarithm, taken as a constant in the present study for simplicity. We have also defined the quantity $\bar{E}_c$, which defines the energy scale above which ion slowing down by electrons becomes dominant, i.e.
\begin{align}
\bar{E}_c &\equiv \frac{m_a}{m_e} \bar{T} \left[ \frac{3\sqrt{\pi}}{4} \left( \frac{n_d}{n_e} \frac{m_e}{m_d} + \frac{n_t}{n_e} \frac{m_e}{m_t} \right) \right]^{2/3}
. \label{eq:ME2sub2}
\end{align}
While $\bar{E}_c \gg 1$ in the hot spot, noting that $\bar{E}_c$ decreases with temperature, this term will impact ion slowing down in the neighboring cold plasma.

\begin{figure}
\begin{centering}
\subfigure[]{\includegraphics[scale=0.5]{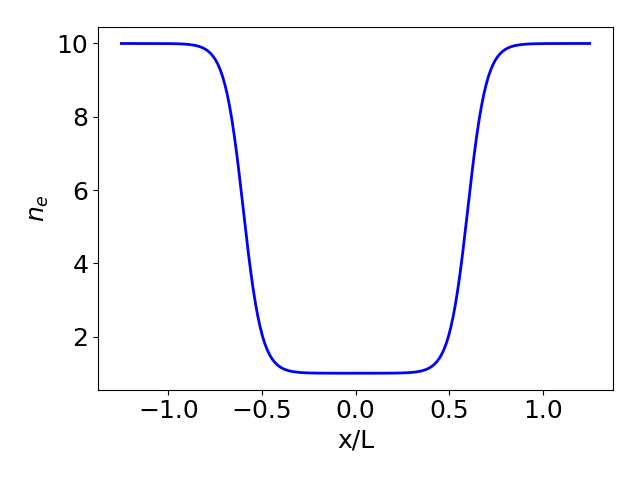}}
\subfigure[]{\includegraphics[scale=0.5]{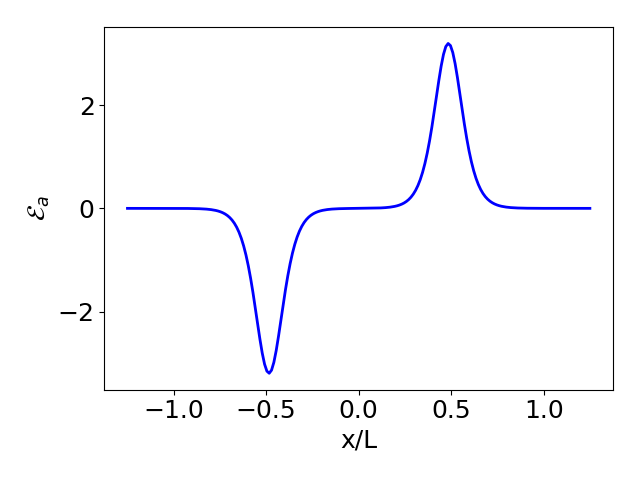}}
\par\end{centering}
\caption{Normalized density [panel (a)] and electric field [panel (b)] profiles used for evaluating the fast ion distribution. The parameters were chosen to be $n_1=9$ and $\Delta x = 0.1$.}
\label{fig:PM0}
\end{figure}

\subsection{\label{sec:HSP}Hot Spot Profiles and Boundary Conditions}

Our aim will be to utilize a PINN to evaluate the ion distribution in an idealized hot spot, where for definiteness we adopt an analogous slab geometry and profiles as used in Ref. \cite{mcdevitt2014comparative}. In particular, we will assume an isobaric equilibrium with density and temperature profiles of the form
\[
\bar{n} = 1 + \frac{n_1}{2} \left[ 2 + \tanh \left( \frac{x_{l} - x}{\Delta x}\right) + \tanh \left( \frac{x - x_{r}}{\Delta x}\right)\right]
,
\]
\[
\bar{T} = 1/\bar{n}
,
\]
where $\bar{n} \equiv n/n^{hs}_e$, $\bar{T} \equiv T/T^{hs}$, and for simplicity we have assumed each particle species to have the same temperature. For all cases treated in this work we will take $x_r=-x_l=0.6$. The electric field will be evaluated from the electron momentum equation, which can be approximated by:
\begin{equation}
\mathcal{E} = \frac{1}{e n_e} \left( -\nabla p_e + R_{ex}\right)
. \label{eq:ME2sub4}
\end{equation}
Here, $R_{ex}$ describes electron-ion momentum exchange, and we have neglected the electron inertia and viscosity terms. For an electron-proton plasma, the thermal force in the momentum exchange term can be written as $R_{ex}=-B_e n_e \nabla T_e$, with $B_e = 0.71$~\cite{braginskii1965reviews}. After normalization, the electric field can then be expressed as:
\begin{equation}
\mathcal{\bar{E}}_a = - 0.71 \frac{e_a}{e} \nabla \bar{T} = - 0.71 \frac{e_a}{e} \frac{\partial \bar{T}}{\partial \bar{x}}
, \label{eq:ME2sub5}
\end{equation}
where since we are considering an isobaric equilibrium, with no equilibrium ion or electron flow, only the thermal force contributes to the electric field. The sign of this electric field is such that it accelerates ions away from the hot spot, thus enhancing fast ion losses. The assumed density and electric field profiles are shown in Fig. \ref{fig:PM0}. For the remainder of this paper we will drop overbars of normalized quantities for notational simplicity unless explicitly indicated.

To complete the description of the problem setup we will need to introduce initial and boundary conditions. With regard to the boundary conditions, we will match the ion distribution to a Maxwellian at the low energy boundary, and take its value to be negligibly small at the high energy boundary. While no rigorous boundary condition is available at the high energy boundary, taking the distribution to a small value was shown in Ref. \cite{mcdevitt2014comparative} to have a minimal impact on the solution of the fast ion distribution across the range of energies that strongly impact the fusion reactivity. The specific value of the distribution at the high energy boundary will be taken to be:
\begin{equation}
f_a \left( E=E_{max} \right) = n_a \left( x\right) \left( \frac{m_a}{2\pi T \left( x\right)} \right)^{3/2} \exp \left(- \frac{E}{T_{min}}\right)
, \label{eq:ME3}
\end{equation}
where $T_{min}$ indicates the minimum temperature in the simulation domain, which in dimensionless units is given by $T_{min} \equiv 1/ \left( 1+n_1\right)$. The initial ion distribution will be taken to be a Maxwellian evaluated at the local density and temperature, but modified to match the high energy boundary condition, i.e.
\begin{equation}
f_a \left( t=0 \right) \equiv f_{a0} = n_a \left( x\right) \left( \frac{m_a}{2\pi T \left( x\right)} \right)^{3/2} \exp \left[- \frac{E}{T \left( x\right)} - \frac{\Delta E^2}{\left( E_{max} - E\right)^2 + \Delta E^2} \left( \frac{E}{T_{min}} - \frac{E}{T \left( x\right)} \right)\right]
, \label{eq:ME4}
\end{equation}
where $\Delta E \ll E_{max}$. For all cases considered in this work we will take $\Delta E = 0.005 E_{max}$. It can be verified that for energies $E \ll E_{max}$ this distribution reduces to a Maxwellian, but at $E=E_{max}$ recovers the high energy boundary condition defined by Eq. (\ref{eq:ME3}). The high energy boundary condition will be taken to be $E_{max}=15$, and the low energy boundary to be $E_{max}=0.01$ for the cases considered in this paper unless explicitly indicated. This high energy boundary condition will enable an accurate solution of the ion distribution for the energies of interest for evaluating the fusion reactivity for hot spots with temperatures of at least two keV. With regard to the low energy boundary, while the test-particle collision operator will not be quantitatively accurate at energies comparable to the thermal energy, the test-particle collision operator described by Eq. (\ref{eq:ME2}) does recover a Maxwellian distribution in the limit of high collisionality.
For an accurate calculation of closure quantities such as the heat flux, plasma viscosity, or momentum and energy exchange rates, which are strongly impacted by first order corrections to the Maxwellian, the low energy boundary should be taken to be several times the thermal energy, and then the tail distribution matched to a particle distribution computed from a collisional closure as described in Ref. \cite{tang2014fusion}. On the spatial boundaries, $x_{min}$ and $x_{max}$, the ion distribution will be taken to be $f_{a0}$.

For the hot spot geometry described above, the fast ion tail is described by the four physics parameters $\left( K^{hs}_n, n_1, \Delta x, n_t/n_d \right)$. Here, $K^{hs}_n$ characterizes the collisionality in the hot spot, $n_1$ and $\Delta x$ quantify the density and temperature difference between the cold and hot regions, together with the steepness of the gradient region, and $n_t/n_d$ is the tritium-deuterium fraction. For the present work we will take $n_t/n_d = 1$ such that we will aim to infer the fast ion solution as a function of the three remaining parameters $\left( K^{hs}_n, n_1, \Delta x \right)$. We note that when evaluating the fusion reactivity, the hot spot temperature $T^{hs}$ will emerge as an important parameter determining the location of the Gamow peak, though it does not appear explicitly in Eq. (\ref{eq:ME2}). This is due to the Chandrasekhar functions being expanded in the limit $v>v_{Ta}$. Its influence thus enters implicitly via the strong dependence of the Knudsen number $K^{hs}_n$ on the hot spot temperature.


\subsection{\label{sec:EPC}Embedding Physical Constraints into the Neural Network}

A key component to ensuring the robust training of the PINN representation of the VFP will be to limit solutions the optimizer searches for to those that are consistent with the physical problem of interest~\cite{arnaud2024physics}. In particular, we will introduce customized input and output layers of the NN that: (i) ensure positivity of $f_a$, (ii) satisfy the low and high energy boundary conditions, (iii) exactly recover $f_{a0}$ as the initial distribution, and (iv) recover known symmetries of the particle distribution. While these constraints could be enforced by a penalty function in the loss of the VFP PINN, by enforcing them as hard constraints this will enable more robust training of the VPF PINN and ultimately a lower loss.

First noting that for the slab description of a hot spot centered about $x=0$ described in Sec. \ref{sec:PM} above, Eq. (\ref{eq:FIS1}) is invariant under the transformation $\left( x, \xi \right) \to \left( -x, -\xi \right)$, indicating that the ion particle distribution must obey
\begin{equation}
f_a \left( -x, -\xi, E, t\right) = f_a \left( x, \xi, E, t\right)
, \label{eq:EPC1}
\end{equation}
This symmetry can be enforced exactly by introducing an additional layer to the NN between the input layer and the hidden layers. In particular, the inputs to the NN will be the independent variables $\left( x, \xi, E, t\right)$ along with the physical parameters $\bm{\lambda}$. The additional layer will take the inputs $\left( x, \xi, E, t; \bm{\lambda} \right)$ and pass them through a layer defined by $\left( x^2, \xi^2, x \xi, E, t; \bm{\lambda} \right)$. Here, the parameter inputs as well as energy and time are simply passed through the additional layer without modification, however, the spatial coordinate and pitch are transformed by $\left( x, \xi \right) \to \left( x^2, \xi^2, x \xi \right)$. Such a transform ensures predictions of the NN satisfy the symmetry indicated by Eq. (\ref{eq:EPC1}).

The additional three constraints indicated above can be enforced by introducing an output layer to the neural network of the form:
\begin{subequations}
\label{eq:IB1}
\begin{equation}
f_a = f_{a0} \hat{f}_a
, \label{eq:IB1a}
\end{equation}
\begin{equation}
\hat{f}_a = \text{exp} \left[ \left( \frac{t}{t_{max}}\right) \left( \frac{E-E_{min}}{E_{max}-E_{min}} \right) \left( \frac{E_{max}-E}{E_{max}-E_{min}} \right) \left( \frac{x-x_{min}}{x_{max}-x_{min}} \right) \left( \frac{x_{max}-x}{x_{max}-x_{min}} \right) \phi_{NN}\right]
, \label{eq:IB1b}
\end{equation}
\end{subequations}
where $f_{a0}$ is the initial ion distribution defined by Eq. (\ref{eq:ME4}) and $\phi_{NN}$ is the output of the hidden layers of the neural network. From Eq. (\ref{eq:IB1}), it is apparent that regardless of the value of $\phi_{NN}$, $f_a$ is: (i) positive definite, (ii) obeys the boundary conditions, and (iii) recovers the initial particle distribution $f_{a0}$ at $t=0$. 


An additional component to the VFP PINN will be the selection of an appropriate form of the loss. Specifically, in addition to the residual of the Vlasov-Fokker-Planck equation, the selection of an appropriate weighting factor will be crucial to ensure the optimizer is able to find an accurate solution across the broad range of energies needed when describing the fast ion distribution. Noting that the boundary and initial conditions are automatically satisfied, and thus do not need to be included in the loss, we will weigh the residual to the VFP equation by the following factors:
\begin{equation}
\text{Loss} = \frac{1}{N_{PDE}} \sum^{N_{PDE}}_i \left[ \left( \frac{1}{\epsilon f^{Max}_{a0} + f_a } \right) \sqrt{ \frac{E_i}{1+E_i} } \mathcal{R}_i \left( x_i, \xi_i, E_i, t_i; \bm{\lambda}_i \right) \right]^2
. \label{eq:IB3}
\end{equation}
Here, $\mathcal{R}_i$ corresponds to the residual of Eq. (\ref{eq:FIS1}), $f^{Max}_{a0} \equiv n_a \left( x\right) \left[ m_a / \left( 2\pi T \left( x\right) \right) \right]^{3/2}$ is a Maxwellian evaluated at $E=0$, and $\epsilon$ is a hyperparameter of the model. The factor $\sqrt{ E/ \left( 1+E \right) }$ softens the divergence of the test-particle collision operator at low energy, but then asymptotes to unity for $E \gg 1$. The factor $1/\left( \epsilon f^{Max}_{a0} + f_a \right)$ serves two purposes. The first is that due to the large density and temperature variation between the hot spot and surrounding cold region, the magnitude of $f_a$ will vary substantially due to $f^{Max}_a \propto n_a/T^{3/2}$. This factor thus helps ensure that the residual is weighted evenly over these regions. In addition, by choosing an appropriate value of $\epsilon$, this factor controls the weighting of different energies. Noting the approximate exponential decay of the ion distribution with energy, a small value of $\epsilon$ leads to the high energy regions of $f_a$ being more heavily weighted, whereas a larger value of $\epsilon$ will give more weight to lower energy regions that have larger values of $f_a$. For all the cases treated in this paper we will take $\epsilon = 0.01$. Further details of the models used in this paper are listed in Table \ref{tab:Params}.

\begin{table}
\footnotesize
\begin{center}
    \begin{tabular}{ | p{2.5cm} | p{2cm} | p{1.6cm} | p{1.6cm} | p{2.0cm} | p{1.6cm} | p{1.6cm} | p{1.6cm} | }
    \hline
    Figures showing results of model & Initial Training points & Time Dependent & Input Transform & $x$ range & $K^{hs}_n$  & $n_1$ & $\Delta x$ \\
    \hline
   
    \hline
    \hline
    Figs. \ref{fig:FIS0}(a), (c), (e)    & $10^6$ & Yes & No  & $\left( -1.25,1.25\right)$ & $\left( 0.01,0.2 \right)$ & $\left( 2,9\right)$ & 0.1 \\
    \hline
    Figs. \ref{fig:FIS0}(b), (d), (f)     & $10^6$ & Yes & Yes & $\left( -1.25,1.25\right)$ & $\left( 0.01,0.2 \right)$ & $\left( 2,9\right)$ & 0.1 \\
    \hline
    Figs. \ref{fig:TE1}, \ref{fig:LM2} & $2\times10^6$ & Yes & Yes & $\left( 0,1.25\right)$ & $\left( 0.01,0.2 \right)$ & $\left( 2,9\right)$ & $\left( 0.05, 0.15\right)$ \\
    \hline
     Figs. \ref{fig:CPR1}, \ref{fig:FIS1}, \ref{fig:FIS2} & $2\times10^6$ & No & Yes & $\left( 0,1.25\right)$ & $\left( 0.01,0.2 \right)$ & $\left( 2,9\right)$ & $\left( 0.05, 0.15\right)$ \\
    \hline
\end{tabular}
\caption{Summary of models used in different figures. All models used a fully connected feedforward neural network with four hidden layers each with a width of sixty-four neurons.}
\label{tab:Params}
\end{center}
\end{table}

\section{\label{sec:FIS}Fast Ion Solution}

\subsection{Impact of Input Transform on the Fast Ion Solution}

\begin{figure}
\begin{centering}
\subfigure[]{\includegraphics[scale=0.5]{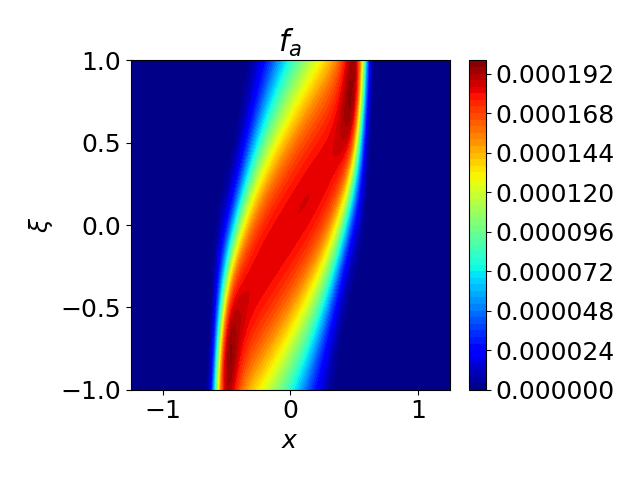}}
\subfigure[]{\includegraphics[scale=0.5]{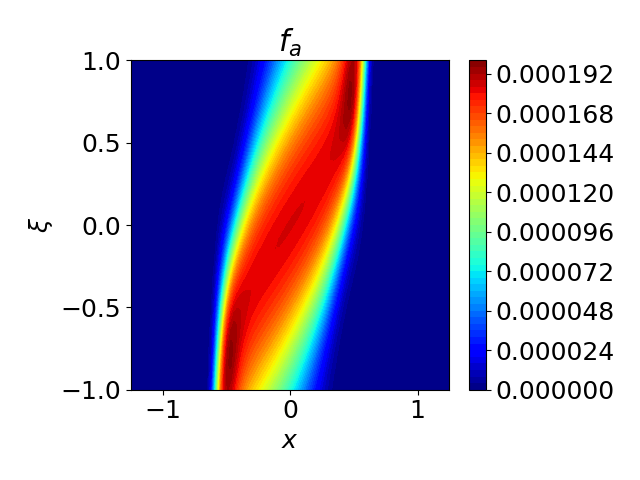}}
\subfigure[]{\includegraphics[scale=0.5]{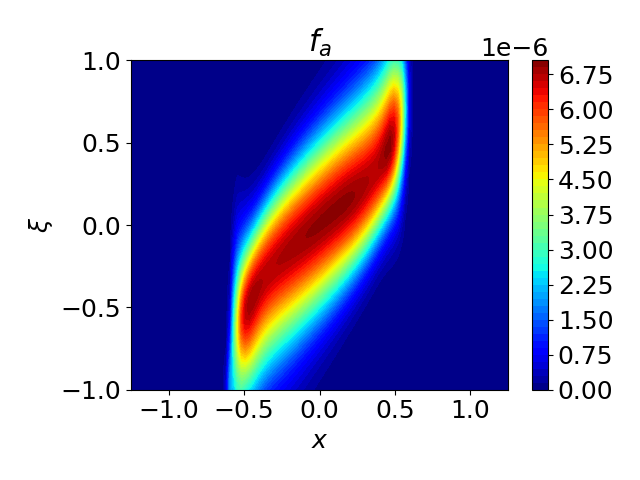}}
\subfigure[]{\includegraphics[scale=0.5]{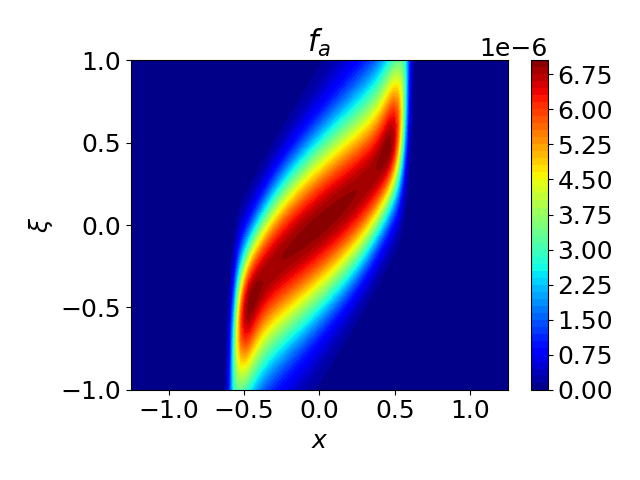}}
\subfigure[]{\includegraphics[scale=0.5]{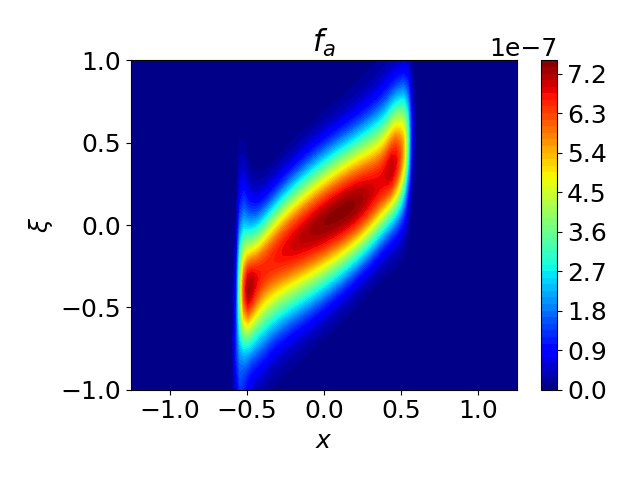}}
\subfigure[]{\includegraphics[scale=0.5]{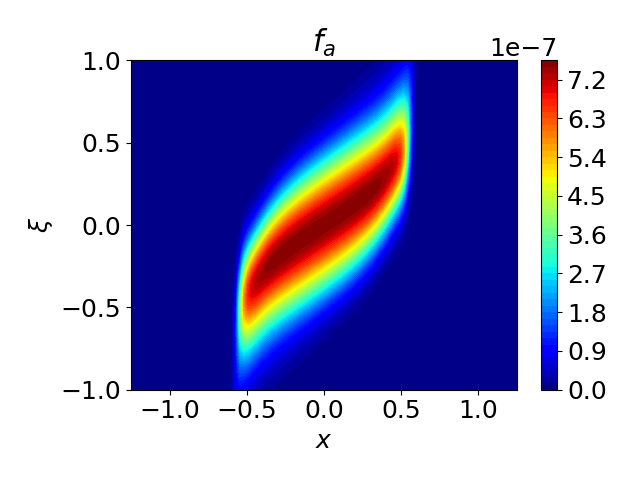}}
\par\end{centering}
\caption{Cross cuts of the fast ion distribution in the $\left( x, \xi \right)$ plane. The cases shown in (a), (c) and (e) do not include the input transform $\left( x, \xi \right) \to \left( x^2, \xi^2, x \xi \right)$ whereas (b), (d) and (f) do. The ion species was chosen to be deuterium, $K_n = 0.15$, $n_1 = 9$, $\Delta x/L = 0.1$ and $t=1$. Panels (a) and (b) are for $E=5$, panels (c) and (d) are for $E=8$, and panels (e) and (f) are for $E=10$.}
\label{fig:FIS0}
\end{figure}

In this section it will be useful to investigate the form of the fast ion solution. We will begin by investigating the impact of the input layer $\left( x, \xi \right) \to \left( x^2, \xi^2, x \xi \right)$ described in Sec. \ref{sec:EPC} above on the fast ion distribution. Figure \ref{fig:FIS0} shows a comparison of the fast ion distribution with and without introducing the input transform (other parameters are indicated in Table \ref{tab:Params}). Considering cross cuts of the fast ion solution in the $\left( x, \xi \right)$ plane at $t=1$ for three different energies, it is apparent that the loss of fast ions from the hot spot leads to substantial deviations from a Maxwellian distribution. In particular, noting that the temperature is maximal in the hot spot between $x \approx \left( -0.6, 0.6\right)$, this is the region which would be expected to have the largest number of fast ions. For the highest energy cross cut shown [$E=10$, Figs. \ref{fig:FIS0}(e) and (f)], it is evident that a large asymmetry in the number of ions with $x \xi > 0$ compared to $x\xi <0$ is present. This is due to the relatively low collisionality at this energy allowing the fast ions to free stream toward the interface between the hot and cold regions. Once these fast ions reach the cold region the high collisionality results in the fast ions slowing down, which leads to maximums of the ion distribution forming at lower energies at $\left( x, \xi \right) \approx \left( 0.6, 1 \right)$ and $\left( x, \xi \right) \approx \left( -0.6, -1 \right)$. These are particularly evident in the $E=5$ cross cut shown in Figs. \ref{fig:FIS0}(a) and (b). While this physics is evident for cases where the input transform is applied [Figs. \ref{fig:FIS0}(b), (d), (f)], and when it is not [Figs. \ref{fig:FIS0}(a), (c), (e)], the latter ion distribution does not precisely satisfy the symmetry $f \left( -x, -\xi, E, t\right) = f \left( x, \xi, E, t\right)$, with the deviations most evident at the highest energy cross cut. 


 \begin{figure}
\begin{centering}
\includegraphics[scale=0.5]{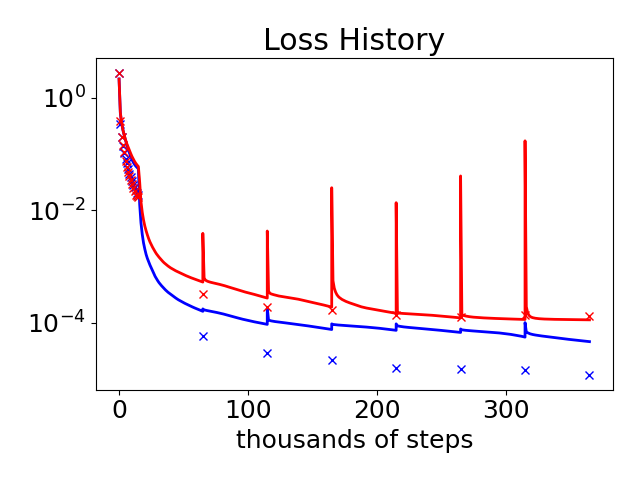}
\par\end{centering}
\caption{Loss history for the ion distributions shown in Fig. \ref{fig:FIS0}. The neural networks contained six inputs $\left( E, \xi, x, t, K^{hs}_n, n_1 \right)$. The solid curves indicate the training loss, whereas the `x' markers indicate the test loss. The blue curves are for the case where an input transform was included, whereas the red curve indicate the case where the input transform was not used.}
\label{fig:FIS0sub2}
\end{figure}

A comparison of the loss histories for the two cases is shown in Fig. \ref{fig:FIS0sub2}. Here an ADAM optimizer is used for the first 15,000 epochs, and L-BFGS is used thereafter. A million training points obeying a Hammersley sequence~\cite{guo2022analysis, wu2023comprehensive} are applied along with 262,144 test points sampled according to a uniform random distribution.  After periods of 50,000 epochs of L-BFGS training, an additional 100 training points are added at locations where the residual is maximal~\cite{lu2021deepxde}, leading to periodic spikes in the training loss. We also note that since the training and test points obey different distributions, we do not expect the magnitude of the training and test losses to match. The addition of training points at locations of maximal residual will further push the training and test losses apart.
It is evident that the loss for the case where an input transform is included is substantially reduced. Hence, the input transform provides a means of reducing the loss, and hence improving the accuracy of the solution, while exactly satisfying a known symmetry of the system. Furthermore, noting the symmetry of the solution is automatically satisfied, it is only necessary to solve for the solution in one half of the spatial domain. Hence for the remainder of this analysis we will limit the spatial domain to positive values of $x$ during training, where the solution for negative values of $x$ can be recovered by noting the symmetry $f \left( -x, -\xi, E, t\right) = f \left( x, \xi, E, t\right)$.

\subsection{\label{sec:TE}Temporal Evolution of the Fast Ion Distribution}

\begin{figure}
\begin{centering}
\includegraphics[scale=0.5]{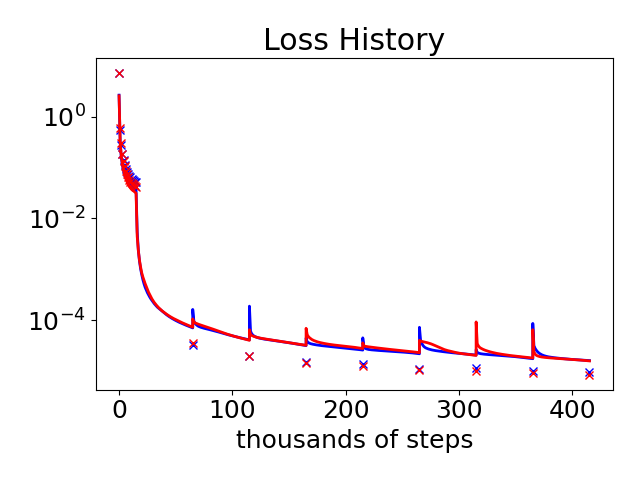}
\par\end{centering}
\caption{Loss histories of PINNs for the tritium and deuterium distributions with seven inputs $\left( E, \xi, x, t, K^{hs}_n, n_1, \Delta x \right)$. The blue curves indicate the loss history for the tritium case, whereas the blue curves indicate the loss for the deuterium case. The solid curves are the training loss, whereas the `x' markers are the test loss.}
\label{fig:TE0}
\end{figure}

\begin{figure}
\begin{centering}
\subfigure[]{\includegraphics[scale=0.5]{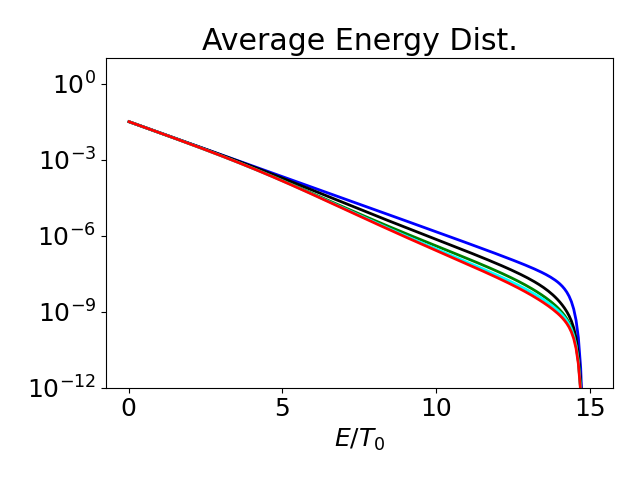}}
\subfigure[]{\includegraphics[scale=0.5]{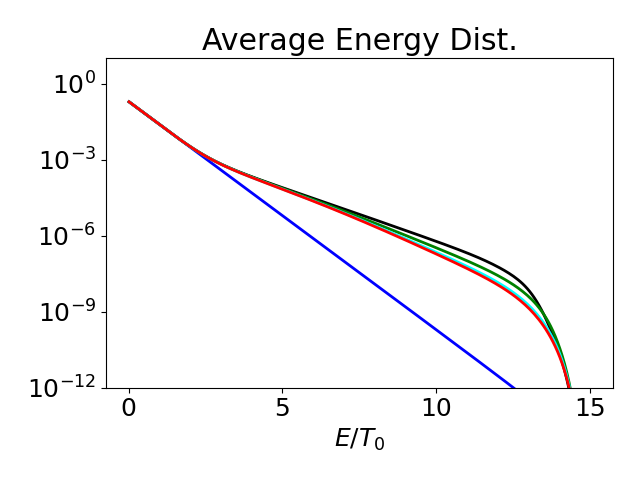}}
\subfigure[]{\includegraphics[scale=0.5]{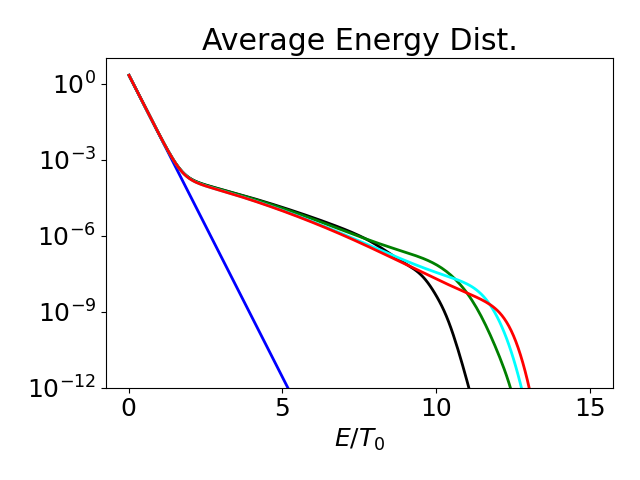}}
\subfigure[]{\includegraphics[scale=0.5]{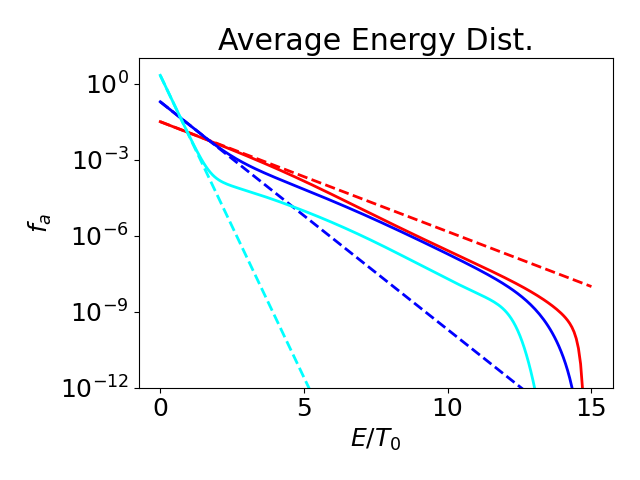}}
\par\end{centering}
\caption{Time slices of the pitch-angle averaged tritium distribution at three different spatial locations. Panel (a) is for $x=0$, panel (b) is for  $x=0.5$ and panel (c) indicates $x=0.6$. The blue curves correspond to $t=0$, the black curves to $t=0.25$, the green curves to $t=0.5$, the cyan curves to $t=0.75$ and the red curves to $t=1$. Panel (d) compares the ion distribution at different spatial locations at $t=1$. The red curves correspond to $x=0$, the blue curves to $x=0.5$, and the cyan curves to $x=0.6$. Dashed curves in panel (d) indicate Maxwellian distributions, whereas solid curves indicate the computed ion distribution. The Knudsen number was taken to be $K_n = 0.15$, $n_1 = 9$, and $\Delta x/L = 0.1$.}
\label{fig:TE1}
\end{figure}

Incorporating the input transform [see the third row of Table \ref{tab:Params} for further details of the model], the loss history for time dependent models of the deuterium and tritium distributions are shown in Fig. \ref{fig:TE0}. The test loss drops by over six orders of magnitude indicating the VFP
PINN was able to successfully train. Time slices of the pitch-angle averaged tritium distribution are
shown at different spatial locations in Fig. \ref{fig:TE1}. Here, Fig. \ref{fig:TE1}(a) indicates five time
slices at $x=0$, where it is evident that while the low energy bulk
plasma remains approximately Maxwellian, the hot tail becomes
significantly depleted by $t=1$. Considering the transition region
[$x=0.5$, Fig. \ref{fig:TE1}(b)], the fast ions lost from the hot spot
lead to an increase in the number of fast ions in the neighboring
region. After this fast ion tail forms, it slowly decreases as the
fast ion tail in the hot spot decreases. Turning to a spatial location
deeper into the cold region [$x=0.6$, Fig. \ref{fig:TE1}(c)], the time
evolution is similar to the $x=0.5$ location, though the fast ions
have slowed down substantially by the time they reach this spatial
location. Fig. \ref{fig:TE1}(d), shows a comparison of the fast ion
distribution at the three spatial locations at $t=1$. From this
comparison it is clear that the magnitude of the fast ion tail
decreases as $x$ is increased, though the slope of the fast ion
distribution with respect to energy is similar at each spatial
location.

\subsection{\label{sec:LM}Legendre Moments of the Fast Ion Distribution}

\begin{figure}
\begin{centering}
\subfigure[]{\includegraphics[scale=0.33]{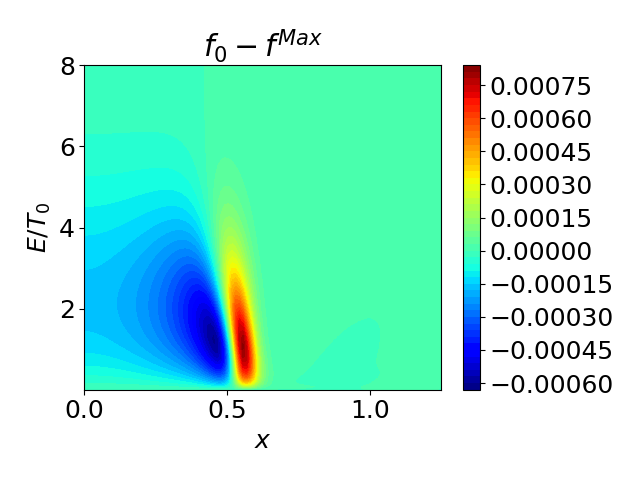}}
\subfigure[]{\includegraphics[scale=0.33]{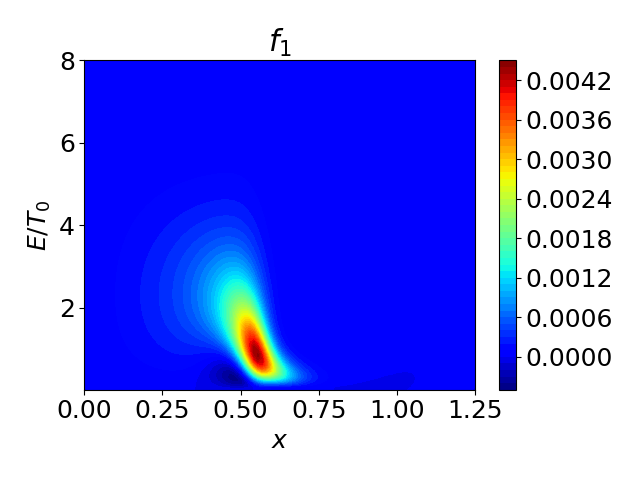}}
\subfigure[]{\includegraphics[scale=0.33]{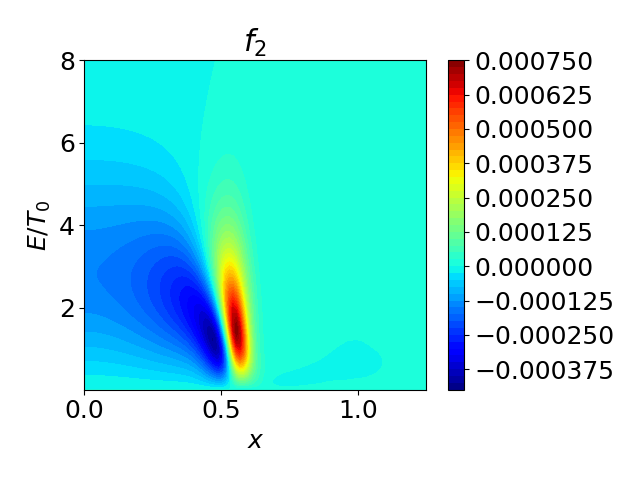}}
\par\end{centering}
\caption{Projections of the tritium distribution onto the first three Legendre polynomials at $t=1$. Panel (a) indicates the difference between $f^{(0)}_t$ and a Maxwellian, panel (b) is the $f^{(1)}_t$ component of the distribution, and panel (c) is the $f^{(2)}_t$ component. The Knudsen number was taken to be $K_n = 0.15$, $n_1 = 9$, and $\Delta x/L = 0.1$.}
\label{fig:LM2}
\end{figure}

More insight into the fast ion solution can be gained by projecting the solution onto a basis of Legendre polynomials $P_l \left( \xi \right)$. In particular, noting the relations
\begin{subequations}
\label{eq:FIS3}
\begin{equation}
\left( n_a , \frac{3}{2} p_a \right) = \int d^3 v \left( 1,\frac{1}{2} m_a v^2 \right) P_0 \left( \xi \right) f_a
, \label{eq:FIS3a}
\end{equation}
\begin{equation}
n_a v_x = \int d^3 v v P_1 \left( \xi \right) f_a
, \label{eq:FIS3b}
\end{equation}
\begin{equation}
p_{a x} - p_{a \perp }  = m_a \int d^3 v v^2 P_2 \left( \xi \right) f_a
. \label{eq:FIS3c}
\end{equation}
\end{subequations}
If the ion distribution is expanded as
\begin{equation}
f_a \left( x,E,\xi \right) = \sum_{l = 0,1,2,\dots} P_l \left( \xi \right) f^{(l)}_a \left( x,E \right)
, \label{eq:FIS4}
\end{equation}
and Eq. (\ref{eq:FIS4}) is substituted into Eq. (\ref{eq:FIS3}), this yields
\begin{subequations}
\label{eq:FIS5}
\begin{equation}
\left( n_a , \frac{3}{2} p_a \right) = 4\pi \int d v v^2 \left( 1,\frac{1}{2} m_a v^2 \right) f^{(0)}_a
, \label{eq:FIS5a}
\end{equation}
\begin{equation}
n_a v_x = \frac{4\pi}{3} \int d v v^3  f^{(1)}_a
, \label{eq:FIS5b}
\end{equation}
\begin{equation}
p_{a x} - p_{a \perp}  = \frac{4\pi}{5} m_a \int d v v^4  f^{(2)}_a
. \label{eq:FIS5c}
\end{equation}
\end{subequations}
It is thus evident that the $f^{(0)}_a$ component describes the density and isotropic pressure, $f^{(1)}_a$ is linked to the spatial flux of fast ions, and $f^{(2)}_a$ describes the pressure anisotropy of the high energy ion tail. The energy and spatial dependence of the first three Legendre coefficients are shown in Fig. \ref{fig:LM2}. Considering the $f^{(0)}_a$ Legendre coefficient, it is evident that the number of fast ions is reduced inside the hot spot, with a substantial surplus present in the neighboring cold region. This depletion of fast ions is mediated by an outflow of fast ions from the hot spot as indicated by a positive value of $f^{(1)}_a$ at high energies.
Considering $f^{(2)}_a$, it is evident that within the hot spot $p_{a x} < p_{a \perp}$ at high energy, indicating that fast ions whose direction is primarily in the $x$ direction are depleted more rapidly than those whose motion is perpendicular to the $x$ direction, with this trend reversed in the neighboring cold region. We note that we do not anticipate the ion solution to be quantitatively accurate at low energies due to the use of a test-particle collision operator expanded in the limit $v>v_{Ta}$. However, as indicated by Fig. \ref{fig:TE1} the model does recover that the ion distribution is nearly Maxwellian near the thermal energy. Noting that the fusion reactivity is most sensitive to the ion distribution at energies of several times the thermal energy, we anticipate that the present model will be most accurate for small to modest Knudsen numbers, where substantial deviations from a Maxwellian distribution only emerge at high energies.


\subsection{\label{sec:CPR}Comparison with Previous Results}

\begin{figure}
\begin{centering}
\includegraphics[scale=0.5]{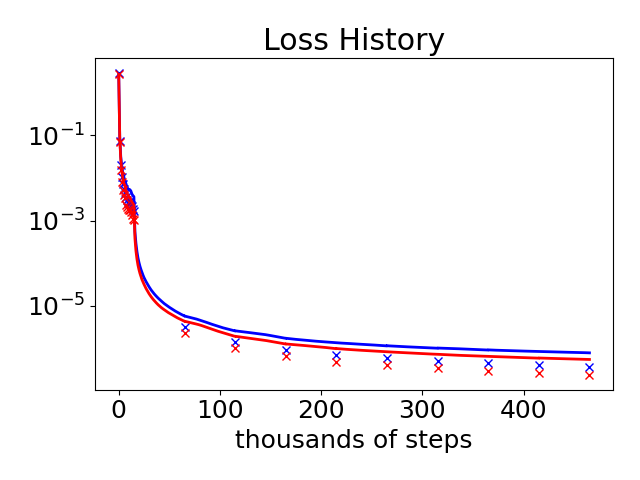}
\par\end{centering}
\caption{Loss histories of steady state VFP PINNs for the tritium and deuterium distributions with six inputs $\left( E, \xi, x, K^{hs}_n, n_1, \Delta x \right)$. The blue curves indicate the loss history for the tritium case, whereas the red curves indicate the loss for the deuterium case. The solid curves are the training loss, whereas the `x' markers are the test loss.}
\label{fig:CPR0}
\end{figure}

Here we will provide a comparison of the deuterium and tritium distribution predicted by the VFP PINN with the results given in Ref. \cite{mcdevitt2014comparative}, which evaluated a nearly identical model of the fast ion distribution using a traditional numerical solver. Noting that Ref. \cite{mcdevitt2014comparative}, considered the limit of a steady state fast ion distribution, we will also modify the VFP PINN to evaluate the steady state fast ion distribution. This is done by removing the time derivative term in Eq. (\ref{eq:FIS1}), and removing time as an input into the VFP PINN. The input parameters for the steady state VFP PINN are thus given by $\left( E, \xi, x, K^{hs}_n, n_1, \Delta x \right)$ [further model details are given in the fourth row of Table \ref{tab:Params}]. The loss history for the steady state VFP PINNs are shown in Fig. \ref{fig:CPR0}. Here, the test loss for both the tritium and deuterium VFP PINNs decreases by nearly seven orders of magnitude implying that the VFP PINNs were able to successfully train.

\begin{figure}
\begin{centering}
\includegraphics[scale=0.5]{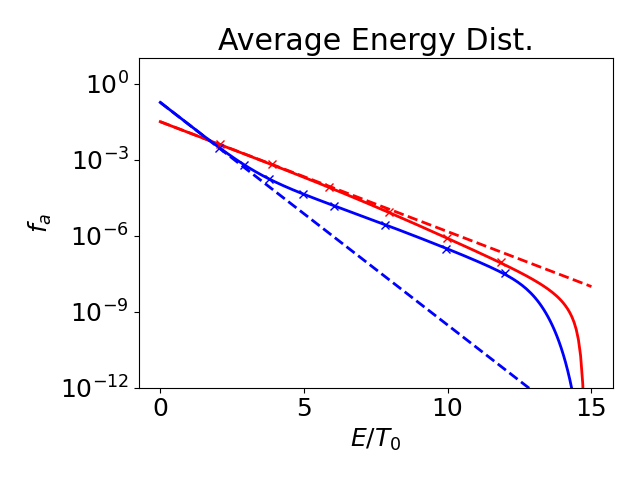}
\par\end{centering}
\caption{Pitch-averaged tritium distribution in the hot spot ($x=0$, red curves) and adjacent cold region ($x=0.4975$, blue curves). The dashed blue curves indicate the nominal Maxwellian distribution, the solid lines indicate predictions from the PINN, and the `x' markers indicate values from Fig. 17 of Ref. \cite{mcdevitt2014comparative} extracted using the software from Ref. \cite{Rohatgi:2017}. The values extracted from Fig. 17 of Ref. \cite{mcdevitt2014comparative} were divided by $2^{3/2}$ to account for the different normalizations used between this paper and Ref. \cite{mcdevitt2014comparative}. The Knudsen number was taken to be $K^{hs}_n = 0.05$, $n_1 = 9$, and $\Delta x/L = 0.1$.}
\label{fig:CPR1}
\end{figure}

A comparison of the predicted pitch-angle averaged tritium
distribution from the VFP PINN and Fig. 17 of
Ref. \cite{mcdevitt2014comparative} is shown in
Fig. \ref{fig:CPR1}. Here, excellent agreement is evident for the ion
distribution in both the hot spot ($x=0$, red curves) and adjacent
cold region ($x\approx 0.5$, blue curves). While the physical
parameters utilized in this comparison were matched, a handful of
numerical parameters differed slightly, yielding modest differences in
the predictions between the present paper and
Ref. \cite{mcdevitt2014comparative}. In particular, the upper and
lower energy bounds employed by the two papers are different. In the
present model, the low energy boundary was taken to be $E_{min} =
0.01$, whereas Ref. \cite{mcdevitt2014comparative} chose $E_{min} =
2$. Furthermore, Ref. \cite{mcdevitt2014comparative} chose $E_{max} =
20$, whereas the present paper selected $E_{max} =
15$. While the PINN implementation could be straightforwardly modified to incorporate a low energy boundary of $E_{min}= 2$, we chose $E_{min} = 0.01$ to demonstrate that PINNs are able to learn the ion distribution across the full range of energies. In contrast, since contributions to the fusion reactivity are negligible beyond $E\approx 15$ for $T^{hs} > 2\;\text{keV}$, we opted to restrict the energy range to $E_{max}=15$.
As a result of these different choices of energy ranges, Fig. \ref{fig:CPR1} only includes values from Fig. 17 of
Ref. \cite{mcdevitt2014comparative} between $E\approx2$ and $E \approx
12$. Furthermore, since Ref. \cite{mcdevitt2014comparative} used a
grid based approach, where the ion distribution was only available at
discreet spatial locations, a careful examination of the Maxwellian
distributions shown in Fig. 17 of Ref. \cite{mcdevitt2014comparative}
indicated that the blue curves in that paper were evaluated at
$x=0.4975$, rather than $x=0.5$ (using the normalization of the
current paper). Finally, while both the present paper and
Ref. \cite{mcdevitt2014comparative} chose a high energy boundary
condition such that the ion distribution was forced to a small value,
this value differed between the two studies yielding slightly
different behavior near the high energy boundary. This choice of
boundary condition, however, does not strongly impact the solution at
energies of interest for evaluating the fusion reactivity for
temperatures greater than a few keV.

\section{\label{sec:KLRR}Knudsen Layer Reactivity Reduction}

In this section we will utilize VFP PINNs for the tritium and deuterium fast ion distributions to quantify how the fusion reactivity varies with hot spot parameters $\left( K^{hs}_n, n_1, \Delta x, T^{hs} \right)$. To accomplish this, the fast ion distributions will be inferred using the steady state VFP PINNs described in Sec. \ref{sec:CPR} above. The fusion reactivity can then be evaluated from the expression
\begin{equation}
\left\langle \sigma v \right\rangle_{ab} \equiv \int d \mathbf{v}_a d \mathbf{v}_b \sigma_{ab} \left( u\right) u f_a \left( \mathbf{v}_a\right) f_b \left( \mathbf{v}_b\right) /n_a n_b
, \label{eq:FR1}
\end{equation}
where $u \equiv \left| \mathbf{v}_a - \mathbf{v}_b \right|$ is the relative velocity and $\sigma_{ab}$ is the fusion cross section parameterized by
\begin{equation}
\sigma_{ab} = \frac{S \left( E_{cm}\right)}{E_{cm}} \exp \left( - \sqrt{\frac{E_G}{E_{cm}} }\right)
. \label{eq:FR2}
\end{equation}
Here $E_{cm}$ is the center-of mass-energy $E_{cm} \equiv m_r u^2/2$, $m_r \equiv m_a m_b / \left( m_a + m_b\right)$ is the reduced mass, $E_G$ is the Gamow energy $E_G \equiv 2 \pi^2 m_r c^2 Z^2_1 Z^2_2 \left( e^2/ \hbar c\right)^2$, and $S \left( E_{cm}\right)$ will be taken to have the form:
\[
S \left( E_{cm}\right) = \frac{A_1 + A_2 E_{cm} + A_3 E^2_{cm} + A_4 E^3_{cm} + A_5 E^4_{cm}}{1 + B_1 E_{cm} + B_2 E^2_{cm} + B_3 E^3_{cm} + B_4 E^4_{cm}}
,
\]
where the specific values of the coefficients $A_i$ and $B_i$ used in the present study will be taken from Ref. \cite{bosch1992improved}. The six-dimensional integral defined by Eq. (\ref{eq:FR1}) can be simplified by expanding in 
Legendre coefficients, yielding \cite{cordey1978new}
\begin{equation}
\left\langle \sigma v \right\rangle_{ab} = \frac{8 \pi^2}{n_a n_b} \sum_{l=0} \frac{1}{2l+1}\int^\infty_0 dv v^2 f^{(l)}_a \left( x,v \right) \int^\infty_0 d v^\prime  {v^\prime }^2 f^{(l)}_b \left(x, v^\prime \right) \int^1_{-1} d \xi_{12} P_l \left( \xi_{12} \right) \sigma_{ab} \left( u\right) u
. \label{eq:FR3}
\end{equation}
Here, the relative velocity can be written as $u \equiv \sqrt{v^2 + {v^\prime}^2 - 2 v v^\prime \xi_{12}}$, the pitch-angle variable is given by $\xi_{12} = \cos \theta_{12}$, where $\theta_{12}$ is the angle between $\mathbf{v}$ and $\mathbf{v}^\prime$, and $f^{(l)}_a$ are the Legendre coefficients defined in Eq. (\ref{eq:FIS4}). As shown in Ref. \cite{mcdevitt2014comparative}, the sum in Eq. (\ref{eq:FR3}) converges rapidly, such that only a few Legendre coefficients are required for convergence. In the present study only the first three Legendre coefficients will be used. The fusion yield can then be written as:
\begin{equation}
Y_{ab} = \epsilon_{ab} R_{ab} \equiv \epsilon_{ab} \frac{n_a n_b}{1+\delta_{ab}} \left\langle \sigma v \right\rangle_{ab}
, \label{eq:FR4}
\end{equation}
where $\epsilon_{ab}$ is the energy released in a given fusion reaction and $R_{ab}$ is the reactivity rate.

\begin{figure}
\begin{centering}
\subfigure[]{\includegraphics[scale=0.5]{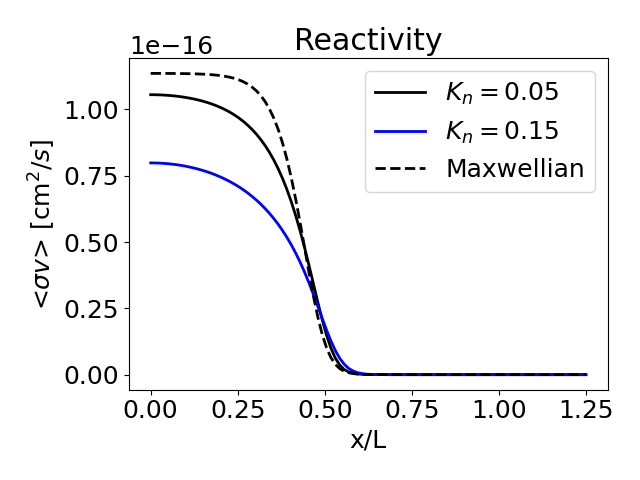}}
\subfigure[]{\includegraphics[scale=0.5]{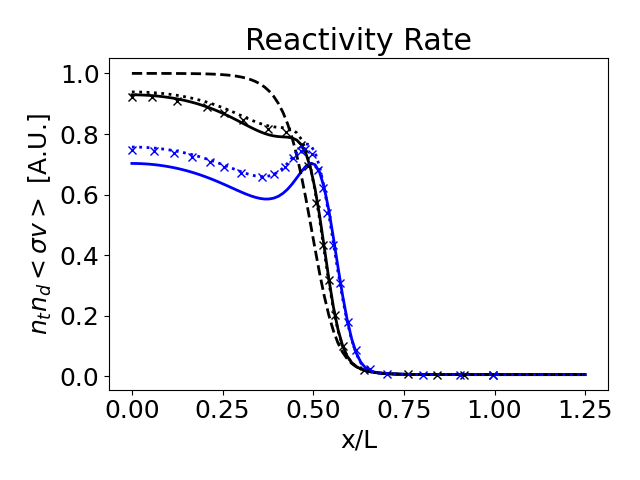}}
\par\end{centering}
\caption{Fusion reactivity [panel (a)] and fusion reactivity rate [panel (b)]. The parameters were taken to be $n_1 = 9$, $\Delta x/L = 0.1$ and $T^{hs} = 10\;\text{keV}$. The `x' markers in panel (b) correspond to values extracted from Fig. 24 of Ref. \cite{mcdevitt2014comparative}, the dotted curves are predictions of a VFP PINN trained with a low energy boundary of $E_{min}=1.5$, whereas the solid curves correspond to the prediction from the VFP PINN with $E_{min}=0.01$. The dashed curves are the nominal Maxwellian reactivity.}
\label{fig:FIS1}
\end{figure}

\begin{figure}
\begin{centering}
\subfigure[]{\includegraphics[scale=0.5]{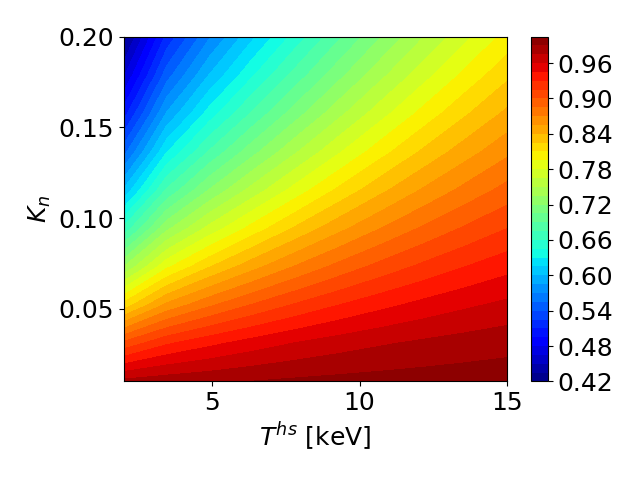}}
\subfigure[]{\includegraphics[scale=0.5]{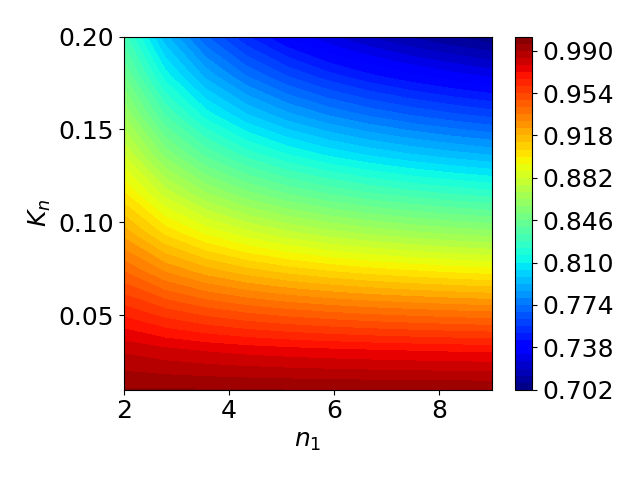}}
\subfigure[]{\includegraphics[scale=0.5]{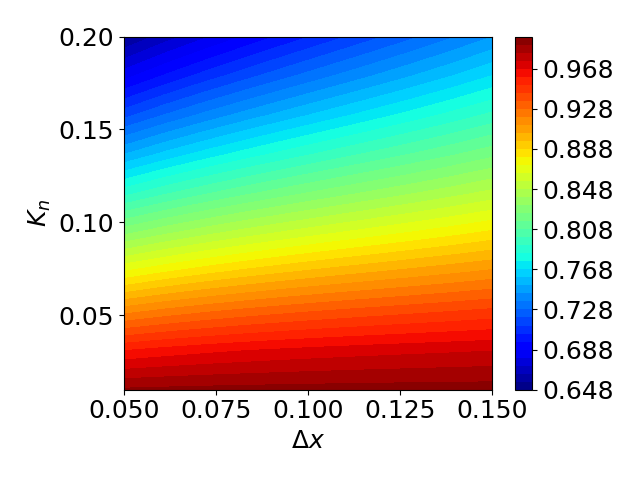}}
\subfigure[]{\includegraphics[scale=0.5]{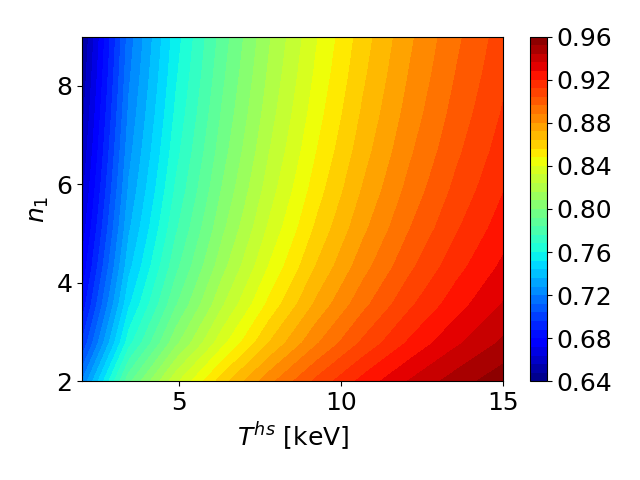}}
\par\end{centering}
\caption{Parametric dependence of spatially integrated fractional yield reduction  $\int^1_0 dx R_{dt} \left( x\right) /  \int^1_0 dx R^{Max}_{dt} \left( x\right)$ versus the hot spot parameters $\left( K^{hs}_n, n_1, T^{hs}, \Delta x \right)$. Panel (a) was evaluated for $n_1=9$ and $\Delta x = 0.1$, panel (b) took $T=10\;\text{keV}$ and $\Delta x = 0.1$, panel (c) was for $n_1 = 9$ and $T^{hs}=10\;\text{keV}$, and panel (d) took $K^{hs}_n=0.1$ and $\Delta x/L = 0.1$.}
\label{fig:FIS2}
\end{figure}

A plot of the reactivity $ \left\langle \sigma v \right\rangle_{dt}$ and reactivity rate $R_{dt}$ for different Knudsen numbers is shown in Fig. \ref{fig:FIS1}. From Fig. \ref{fig:FIS1}(a), it is evident that as the Knudsen number is increased the fusion reactivity is substantially decreased in the hot spot. The escaping fast ions, do however, introduce a slight increase in the reactivity in the neighboring cold region. While this increase in reactivity is small, the reaction rate $R_{dt} = n_d n_t \left\langle \sigma v \right\rangle_{dt}$ is substantially increased due to the high density present in the neighboring cold region [see Fig. \ref{fig:FIS1}(b)]. Despite this increase in the neighboring cold region, the net fusion yield ($\int^1_0 dx R_{dt}$) decreases as $K^{hs}_n$ increases [see Fig. \ref{fig:FIS2} below], suggesting radiation-hydrodynamic codes that are based on the assumption of a Maxwellian plasma will overpredict the fusion reactivity at high temperatures and low densities where the Knudsen number is largest.

A comparison of the spatial fusion reaction rate profile with Fig. 24 of Ref. \cite{mcdevitt2014comparative} is shown in Fig. \ref{fig:FIS1}(b). Here the solid curves represent the predictions of the VFP PINN whereas the `x' markers indicate values shown in Fig. 24 of \cite{mcdevitt2014comparative} extracted using the software from Ref. \cite{Rohatgi:2017}. Due to the different normalizations employed between the present work and Ref. \cite{mcdevitt2014comparative}, both datasets have been normalized to the fusion reactivity at the center of the hot spot for a Maxwellian ion distribution. Considering first the case of a modest Knudsen number (solid black curve, $K^{hs}_n=0.05$) the VFP PINN is in excellent agreement in the region adjacent to the hot spot, with good agreement also evident inside the hot spot. In contrast, for a large Knudsen number of $K^{hs}_n=0.15$, the VFP PINN predicts lower fusion yield across the entire hot spot. The origin of this systematic shift in predictions is that the low energy boundary conditions were different for the two cases. In particular, Ref. \cite{mcdevitt2014comparative} matched the fast ion solution to a Maxwellian at $E/T^{hs} = 2$ for the tritium distribution, and $E/T^{hs} = 4/3$ for the deuterium distribution. In the present work we have taken $E/T^{hs} = 0.01$ for both ion distributions. As the Knudsen number is increased, deviations from a Maxwellian distribution will emerge at lower energies, which were not entirely captured by Ref. \cite{mcdevitt2014comparative} due to the low energy boundary conditions applied in that work. In contrast, while the present study evaluates nearly the entire distribution, the fast ion model employed is not quantitatively accurate for $E/T \lesssim 1$. Hence, if deviations from a Maxwellian emerge at $E/T \sim 1$, these will not be accurately quantified by the collision coefficients defined by Eq. (\ref{eq:ME2}). As discussed in Ref. \cite{mcdevitt2014comparative}, the fast ion model employed will thus be most accurate for cases of small to modest Knudsen numbers.

The dependence of the fusion reactivity rate on the low energy boundary is confirmed by training a VFP PINN with a low energy boundary of $E_{min}=1.5$ for both tritium and deuterium. The results are shown by the dotted curves in Fig. \ref{fig:FIS1}(b). Here it is evident that for $K^{hs}_n = 0.05$ the reactivity predictions are in good agreement for all three cases. However, for the case with $K^{hs}_n = 0.15$ the location of the low energy boundary significantly impacts predictions of the fusion reactivity. In particular, for the case with $E_{min}=1.5$ the VFP PINN is now in good agreement with Ref. \cite{mcdevitt2014comparative}, which employed a similar low energy boundary. This sensitivity to the low energy boundary for $K^{hs}_n = 0.15$ implies that the test-particle VFP equation will not give quantitatively accurate predictions for Knudsen numbers of this magnitude, though it does recover the qualitative trend of reduced reactivity in the center of the hot spot, with a subsequent increase in the adjacent region.

More insight into how the hot spot parameters impact Knudsen layer reactivity reduction can be gained by considering the dependence of the net fusion yield on $\left( K^{hs}_n, n_1, \Delta x, T^{hs} \right)$. Here, we will evaluate the spatially integrated fusion yield $\int^1_0 dx R_{dt} \left( x\right)$ as inferred from the VFP PINN, and then compare with the nominal Maxwellian value $\int^1_0 dx R^{Max}_{dt} \left( x\right)$. The ratio of these two quantities is shown in Fig. \ref{fig:FIS2}. First considering the dependence of the Knudsen number $K^{hs}_n$, it is apparent that the spatially integrated fusion yield always decreases with increasing $K^{hs}_n$. This is due to larger Knudsen numbers enabling more fast ions to escape the hot spot, where they will collide with colder ions and hence will on average have a lower relative velocity, and thus lower fusion cross section. Furthermore, fast ions that escape from the hot spot will lose some of their energy to ion-electron collisions, further reducing the net fusion yield. We caution that while we are showing results across the full range of Knudsen numbers included in the training of the VFP PINN, the fast ion model will be most accurate for small to modest Knudsen numbers as discussed above.

Turning to the dependence on the hot spot temperature $T^{hs}$, the ratio $\int^1_0 dx R_{dt} \left( x\right) / \int^1_0 dx R^{Max}_{dt} \left( x\right)$ increases as the temperature is increased. This is due to the location of the Gamow peak normalized to temperature decreasing as $T^{hs}$ is increased. Specifically, for ions obeying a Maxwellian distribution, the energy of the Gamow peak normalized to the local temperature is given by:
\[
\frac{E^{GP}_{cm}}{T} = \left( \frac{E_G}{4T} \right)^{1/3}
,
\]
which decreases as $1/T^{1/3}$. Since the depletion of the ion distribution at low values of $E/T$ is less pronounced compared to higher energies,  the fractional reduction of fusion yield is also reduced. From Figs. \ref{fig:FIS2}(b) and (c) it is also evident that the Knudsen layer yield reduction becomes more severe as the temperature and density change between the hot spot and cold region become more extreme, or as the density gradient between these regions becomes sharper. This is due to the density and temperature variation providing the drive for fast ion losses, such that as $n_1$ is increased, or the gradient length scale $\Delta x$ is reduced, larger reductions on the fusion yield are expected.

\section{\label{sec:CD}Conclusions and Discussion}

A physics-informed neural network was used to evaluate the fast ion
tail of the tritium and deuterium distribution in the context of
Knudsen reactivity layer depletion. This approach was shown to yield
accurate predictions of the fast ion distribution in both the hot spot
and neighboring cold region, and thus provide a robust description of
fast ion depletion. A feature of the present approach is that the
offline training time for each network was long, requiring
approximately a day to reach a saturated level of loss on an Nvidia
A100 GPU. However, the online inference time is short, typically only
a microsecond per prediction. While this offline training time is
expected to be long compared to most traditional fast ion solvers, a
single VFP PINN is able to learn the parametric dependence $\left( K^{hs}_n, n_1, \Delta x \right)$ of the fast ion solution, thus the model only needs to be trained once, and can then be deployed to efficiently
explore the parameter space. This can be contrasted with a traditional
fast ion solver, where the run time for a geometry of this complexity
would be expected to be far shorter than the training time of the VFP
PINN, but the traditional solver would need to be rerun for each
parameter set. While the present model involves a relatively low 3-D parameter space, we do not anticipate the treatment of more comprehensive descriptions of the target hot spot (including a variable mix of materials, for example) to pose a fundamental obstacle. Such a study will be carried out in future work.

While in the present paper our focus was on Knudsen layer reactivity reduction, with appropriate modifications to the VFP PINN, additional quantities of interest linked to a non-Maxwellian tail distribution can be evaluated analogously. Planned improvements to the VFP PINN include the incorporation of the field-particle collision operator, which is essential for evaluating closure quantities such as the heat flux. Furthermore, we do not anticipate any difficulties in extending the approach to a fast electron population. We thus expect the present approach to provide a path through which surrogate models can be developed for a range of plasma kinetic effects. The exploration of this approach for the purpose of providing a non-perturbative evaluation of the fast particle distribution for a broader range of applications will be left to future work.

\begin{acknowledgements}

This work was supported by DOE OFES under award number DE-SC0024634. The authors acknowledge the University of Florida Research Computing for providing computational resources that have contributed to the research results reported in this publication.

\end{acknowledgements}

\newpage

%
%
%
%

\bibliographystyle{apsrev}
\bibliography{./ref}

\end{document}